# Accurate Quark CKM Mixing Matrix in Terms of One Universal Parameter

E. M. Lipmanov
40 Wallingford Road # 272, Brighton MA 02135, USA

**Abstract**
The quark Cabibbo-Kobayashi-Maskawa mixing matrix is a fundamental part of the Standard Model accurately determined to fit world data analysis. In this paper the CKM matrix elements are high accurately rendered via simple compact parameterization by one small dimensionless parameter $\varepsilon$. Unique value of the KM CP-violating phase $\delta_q = -65.53°$ is derived in agreement with fitting the data Wolfenstein parameterization. With the three mixing angles $\theta_c$ (Cabibbo angle), $\theta_{23}$, $\theta_{13}$ and CP-violating phase $\delta_q$ quantitatively determined, the complete content of the CKM global fit in the SM is reproduced well within the small relative 1 S.D. ranges in the form of a united system. $\varepsilon$-Parameterization of the neutrino Pontecorvo-Maki-Nakagava-Sakata mixing matrix in agreement with available data is related to the quark CKM one by the idea of quark-neutrino mixing angle complementarity extended to the relation between Dirac CP-violating phases $\delta_\ell$ and $\delta_q$.

## 1. $\varepsilon$-Parameterization of the quark mixing matrix

The standard parameterization of the unitary Cabibbo-Kobayashi-Maskawa quark mixing matrix [1], [2] is in the form

$$V \cong \begin{pmatrix} C_{12}\,C_{13} & S_{12}\,C_{13} & S_{13}\,e^{-i\delta} \\ -S_{12}C_{23}-C_{12}\,S_{23}S_{13}\,e^{i\delta} & C_{12}C_{23}-S_{12}\,S_{23}S_{13}\,e^{i\delta} & S_{23}C_{13} \\ S_{12}S_{23}-C_{12}\,C_{23}S_{13}\,e^{i\delta} & -C_{12}S_{23}-S_{12}\,C_{23}S_{13}\,e^{i\delta} & C_{23}C_{13} \end{pmatrix}, \quad (1)$$

where three mixing angles are introduced by

$$S_{12} \equiv \sin\theta_c,\ S_{23} \equiv \sin\theta_{23},\ S_{13} \equiv \sin\theta_{13}, \quad (2)$$

$C_{ij} \equiv \cos\theta_{ij}$, $\theta_c$ is the Cabibbo angle and $\delta$ is the Kobayashi-Maskawa [2] phase responsible for all CP-violating phenomena in quark flavor-changing processes in the SM.

In this paper we observe that the complete highly accurate content [5] of the quark CKM mixing matrix can be reproduced in terms of one empirical universal parameter (denoted $\varepsilon$).

The goal is achieved by $\varepsilon$-parameterization of the three quark mixing angles through sin-double-angle exponential relations

$$\sin^2(2\theta_c) \cong (2\varepsilon)\exp(2\varepsilon), \quad (3)$$

$$\sin^2(2\theta_{23}) \cong (\varepsilon^2)\exp(\varepsilon^2), \quad (4)$$

$$\sin^2(2\theta_{13}) \cong (\varepsilon^4)\exp(2\varepsilon). \quad (5)$$

Note that at leading $\varepsilon$-approximation the $\varepsilon$-power pattern in (3)-(5) is a semi geometric one

$$\{\sin^2(2\theta_c),\ \sin^2(2\theta_{23}),\ \sin^2(2\theta_{13})\} \cong \{2\varepsilon,\ \varepsilon^2,\ \varepsilon^4\};$$

and the values of the two largest quark mixing angles are described by two conformal exponential connections $\sin^2(2\theta_{1,2}) = (x_{1,2})\exp(x_{1,2})$ with $x_1 = 2\varepsilon$ and $x_2 = \varepsilon^2$ for $\theta_1=\theta_c$ and $\theta_2=\theta_{23}$ respectively.

The magnitude of the parameter $\varepsilon$ can be accurately estimated from experimental data of one mixing angle, e.g. the Cabibbo one $\tan\theta_c = V_{12}/V_{11}$. With exception of the $\delta$- phase, all elements of the matrix (1) can then be expressed through the ε–parameter by relations (3)–(5).

Below we take advantage of another start that emphasized an interesting unity of particle flavor quantities such as lepton and quark mass ratios, mixing angles at cetera.

Earlier a universal empirical flavor-electroweak parameter [3] $\varepsilon \cong 0.082$ is introduced from analysis of experimental data of lepton and quark mass ratios, mixing matrices and precise fitting to the fine structure constant at zero momentum transfer $\alpha \equiv \alpha(q^2 = 0)$,

$$(\exp \alpha / \alpha)^{\exp 2\alpha} + (\alpha/\pi) = 1/\varepsilon^2. \qquad (6)$$

With $\varepsilon^2 = e^{-5}$ this relation is accurate to within $10^{-8}$.

Complete high accurate content of the CKM quark mixing matrix, which is a world average fit to available experimental data [5], can be precisely rendered in terms of one universal ε-parameter

$$\varepsilon = \exp(-5/2). \qquad (7)$$

The three angles $\theta_c$, $\theta_{23}$ and $\theta_{13}$ are accurately expressed through this parameter by the relations (3)-(5):

$$\theta_c \cong 13.047°, \quad \theta_{23} \cong 2.362°, \quad \theta_{13} \cong 0.21°. \qquad (8)$$

As final result, the matrix (1) with the three angles $\theta_c$, $\theta_{23}$ and $\theta_{13}$ from (8) leads to a quantitative prediction of the unitary quark mixing matrix given by

$$V \cong \begin{pmatrix} 0.97418 & 0.22575 & 0.00366\, e^{-i\delta} \\ -0.2256 - 0.00015\, e^{i\delta} & 0.97336 - 0.00003\, e^{i\delta} & 0.0412 \\ 0.00930 - 0.0036\, e^{i\delta} & -0.0402 - 0.0008\, e^{i\delta} & 0.999143 \end{pmatrix}. \qquad (9)$$

From this detailed prediction follows a matrix of absolute matrix elements

$$V_a \cong \begin{pmatrix} 0.97418 & 0.22575 & 0.00366 \\ 0.2256 & 0.97336 & 0.0412 \\ 0.00847 & 0.0405 & 0.999143 \end{pmatrix}. \quad (9')$$

It should be compared with the PDG [5] quantitative presentation of the global fit in the Standard Model

$$V_{CKM} \cong \begin{pmatrix} 0.97419 \pm 0.00022 & 0.2257 \pm 0.0010 & 0.00359 \pm 0.00016 \\ 0.2256 \pm 0.0010 & 0.97334 \pm 0.00023 & 0.0415 \pm 0.0010 \\ 0.00874 \pm 0.0003 & 0.0407 \pm 0.0010 & 0.999133 \pm 0.000044 \end{pmatrix}. \quad (10)$$

The agreement between the obtained matrix (9) and the data one (10) is excellent, always well within the small relative 1 S.D. ranges.

As a result, high accurate values of all three CKM mixing angles are derived in a united form (3)-(5) and (8), which is structured by one empirical universal parameter without taking into account SM radiative correction. Assuming that that kind of agreement gets confirmed by further more accurate experimental tests, it is difficult to envisage what kind of necessary substantially new physics beyond the SM controls this accurate phenomenology.

The CP-violating phase $\delta$ is derived in [6] as a solution of an equation that follows from the discussed quite general quadratic deviation-from-mass-degeneracy hierarchy rule in flavor phenomenology. A more accurate defining equation for the CP-phase is chosen here in the form

$$\sin^2 \delta = 2 \cos \delta. \quad (11)$$

It has only two solutions[1]

$$\delta_q \cong \pm 65.53°. \quad (12)$$

---

[1] Note that this solution obeys an approximate relation $\cos^2 \delta_q \cong (2\varepsilon) \exp(\varepsilon/2)$ accurate to within $6 \times 10^{-4}$.

In the Wolfenstein parameterization [7] the CKM fit [5] to experimental data is given by

$$V \cong \begin{pmatrix} 1 - \lambda^2/2 & \lambda & A\lambda^3(\rho - i\eta) \\ -\lambda & 1 - \lambda^2/2 & A\lambda^2 \\ A\lambda^3(1-\rho - i\eta) & -A\lambda^2 & 1 \end{pmatrix} + O(\lambda^4), \quad (13)$$

the four parameters are:

$$\lambda \cong 0.2257 \pm 0.0010, \quad A \cong 0.814 \pm 0.022, \quad \rho \cong 0.135 + 0.031 - 0.016,$$
$$\eta \cong 0.349 + 0.015 - 0.017.. \quad (14)$$

By comparing the matrices (9) and (13), values of the four Wolfenstein parameters are predicted

$$\lambda \cong 0.22575, \quad A \cong 0.80843, \quad \rho \cong 0.1630, \quad \eta \cong 0.35816,$$
$$\delta_q \cong -65.53°. \quad (15)$$

They are in good agreement with data values (14), always within the 1 S.D. ranges. So, the quadratic-hierarchy equation (11) for the CP-violation phase in the CKM matrix has a unique and fitting solution and is interesting.

The obtained quark mixing matrix (9) determines the CP-violating parameter $\beta \equiv \beta_{\psi K}$

$$\sin 2\beta \cong 0.713 \quad (16)$$

where $\beta = \arg(-V_{cd}V_{cb}/V_{td}V_{tb})$. It agrees with the experimental BABAR and Belle data analysis from [5]

$$\sin 2\beta = 0.681 \pm 0.025 \quad (17)$$

to within 1.3 S.D.

The Jarlskog [8] invariant of the obtained quark mixing matrix (9) with the CP-violating phase (15) is

$$J_q = \text{Im}(V_{12} V_{23} V^*_{13} V^*_{22}) \cong 3.016 \times 10^{-5}. \quad (18)$$

## 2. ε-Parameterization of the neutrino mixing matrix

To parameterize the Pontecorvo-Maki-Nakagawa-Sakata neutrino mixing matrix [9] in terms of the universal ε-parameter we use the observation of quark-neutrino mixing angle complementarity [10] for the two largest mixing angles $\theta_{12}$ and $\theta_{23}$. It means that the equations for those two neutrino angles should be obtained from the quark ones (3) and (4) just by replacements

$$(\sin^2 2\theta_c)_q \to (\cos^2 2\theta_{12})_\ell, \quad (\sin^2 2\theta_{23})_q \to (\cos^2 2\theta_{23})_\ell. \qquad (19)$$

The experimental data of the neutrino mixing matrix are still much less accurate than the CKM quark ones. Therefore the parameterization of the PMNS mixing matrix can be simplified in comparison with the quark case, without considering exponential factors[2]. And so, the two largest neutrino mixing angles are determined by relations

$$\cos^2 2\theta_{12} \cong 2\varepsilon, \quad \theta_{12} \cong 33.05^\circ, \qquad (20)$$

$$\cos^2 2\theta_{23} \cong \varepsilon^2, \quad \theta_{23} \cong 42.65^\circ. \qquad (21)$$

They are connected by the quadratic hierarchy rule: $\cos^2 2\theta_{12} \cong 2\cos 2\theta_{23}$, comp.[4].

We extend the complementarity condition to the neutrino and quark CP-violating phases $\delta_\ell$ and $\delta_q$. With the quark equation (11), the defining equation for the neutrino Dirac CP-violating phase is given by

$$\cos^2 \delta_\ell = 2 \sin \delta_\ell, \quad \delta_\ell + \delta_q = 90^\circ. \qquad (22)$$

Its unique solution is

$$\delta_\ell \cong 155.53^\circ. \qquad (23)$$

Since the general form of unitary neutrino mixing matrix in standard parameterization is again given by (1), the one remaining not defined neutrino mixing angle is $\theta_{13}$.

---

[2] Chronologically it was a reverse order. First (2007) were found the fittings to the neutrino solar and atmospheric mixing angle data in quadratic double-angle forms (20) and (21), and then they were extended to quark mixing.

From the neutrino data the reactor mixing angle is small and cannot be related to the corresponding quark small angle (5) by the complementarity condition. The quark and neutrino (1-3) mixing angles, in contrast to the generic pairs $(\theta_c, \theta_{23})_q$ and $(\theta_{12}, \theta_{23})_\ell$, are two 'singles'. With the presumption that between the quark and neutrino mixing matrices must be a connection, an interesting possibility is that the two small quark and neutrino mixing angles are another generic pair obeying the quadratic hierarchy rule (at leading $\varepsilon$-approximation)

$$\sin^2 2\theta^\ell_{13} = 2\sin 2\theta^q_{13}. \tag{24}$$

With the quark value (5) we get

$$\sin^2 2\theta^\ell_{13} \cong 2\varepsilon^2, \quad \theta^\ell_{13} \cong 3.4°. \tag{25}$$

Finally, the predicted neutrino mixing matrix with Dirac phase (23) is given by

$$V_\nu \cong \begin{pmatrix} 0.837 & 0.544 & 0.06\,e^{-i\delta} \\ -0.4 - 0.0345\,e^{i\delta} & 0.617 - 0.0224\,e^{i\delta} & 0.676 \\ 0.37 - 0.037\,e^{i\delta} & -0.568 - 0.0244\,e^{i\delta} & 0.734 \end{pmatrix}. \tag{26}$$

The Jarlskog [8] invariant of the neutrino mixing matrix (26) with the CP-violating phase (23) is

$$J_\ell \cong 5.6 \times 10^{-3}, \tag{27}$$

It is more than two orders of magnitude larger than the quark one (18).

The neutrino mixing matrix (26) is in good agreement with the analysis of available experimental data [11], to within few percent from the central values. The deviation from the tri-bimaximal mixing [12] is in the range of a few percent; the atmospheric mixing angle is large, but definitely deviated from maximal.

Some noticeable approximate relations between different neutrino mixing angles and the Dirac CP-violating phase are

$$\sin 2\theta^{\ell}{}_{13} \cong \sqrt{2}\cos 2\theta_{23} \cong \sin^2 \delta_{\ell}/\sqrt{2}. \qquad (28)$$

Note that from (28) maximal atmospheric mixing in neutrino oscillations would mean no lepton Dirac CP-violation affects.

From (25), the CP-violating parameter $\sin^2 2\theta^{\ell}{}_{13} \cong 0.013$ is compatible with the obtained in ref. [13] limits from global data analysis,

$$(\sin^2 \theta^{\ell}{}_{13})^{\exp} \cong 0.016 \pm 0.010\,(1.6\,\sigma). \qquad (29)$$

The result from (25) $\sin^2 2\theta^{\ell}{}_{13} \geq 0.01$, though small, is interesting since it supports the hopes [14] of observing lepton CP-violation in the super-beam neutrino experiments.

### 3. Conclusion

The quark Cabibbo-Kobayashi-Maskawa mixing matrix is a fundamental part of the Standard Model, its accurate presentation in form of a closely united system should be interesting. In this paper the world fit CKM matrix in the SM is accurately re-constructed in terms of one universal ε-parameter with no SM radiative corrections. The 9 accurate quark mixing matrix elements appear a closely united system, which is reasonably related to particle mass ratios and fine structure constant via the ε-parameter, comp. [4]. Accurate phenomenology of the CKM mixing matrix points to new basic physics.

Unique and fitting data values of the Dirac CP-violation phases $\delta_q = -65.53°$ and $\delta_{\ell} = 155.53°$ in quark and neutrino mixing matrices are reassuring.

High accurate agreements of ε-parameterized elements of the matrix (9) with the world fit matrix (10), and (12),

are quantitative empirical physical facts and thus are new manifestations of the universal nature of the ε-parameter.

Substantial relation between quark and neutrino mixing matrices is in the spirit of the SM. In this paper ε-parameterization of the neutrino PMNS mixing matrix is achieved in quantitative agreement with available data by the idea of quark-neutrino mixing angle complementarity extended to the Dirac CP-violating phases.

### Appendix. On relation of the ε-parameter to GUT physics

In contrast to the main text, this Appendix is rather a speculative comment. A partial answer to the question of what is the source of the ε-parameter's ubiquity, especially its connections to the low energy electroweak interaction constants, is that it is related to the unique interaction constant $\alpha_{GUT}$ of Grand Unification Theory. A tentative relation that is probably fitting the RGE estimations of connection between the low energy fine structure constant and its GUT value $\alpha_{GUT} = (g_{GUT}^2/4\pi)$ is given by

$$\varepsilon \cong 2\alpha_{GUT}, \quad \alpha_{GUT} \cong \varepsilon/2 \cong 1/24. \quad (A1)$$

Note that the hierarchy of the two extreme values of the fine structure constant, at $q^2=0$ and $q^2=q_{GUT}^2$, appears in accord with the quadratic hierarchy paradigm in flavor phenomenology

$$\alpha_{q^2=0} \approx 4\alpha_{GUT}^2. \quad (A2)$$